\documentclass{article}

\usepackage{arxiv}
\newcommand{\hide}[1]{}
\usepackage[utf8]{inputenc} 
\usepackage[T1]{fontenc}    
\usepackage{hyperref}  
\usepackage{wrapfig}
\usepackage{graphicx}
\usepackage[export]{adjustbox}
\usepackage{xcolor}
\newcommand{\foo}{\hspace{-2.3pt}\textcolor{blue}{$\bullet$} \hspace{5pt}}
\usepackage{subfigure}
\usepackage{paralist}
\usepackage{url}            
\usepackage{booktabs}       
\usepackage{amsfonts}       
\usepackage{nicefrac}       
\usepackage{microtype}      
\usepackage{lipsum}	
\title{Voice for the Voiceless: Active Sampling to Detect Comments Supporting the Rohingyas}

\author{
  Shriphani Palakodety$^*$\\
  Onai\\
  \texttt{spalakod@onai.com} \\
   \And
Ashiqur R. KhudaBukhsh\thanks{Shriphani Palakodety and Ashiqur R. KhudaBukhsh are equal contribution first authors. Ashiqur R. KhudaBukhsh is the corresponding author.} \\
  Carnegie Mellon University \\
  \texttt{akhudabu@cs.cmu.edu} \\
  \And
Jaime G. Carbonell \\
  Carnegie Mellon University\\
  \texttt{jgc@cs.cmu.edu} \\
}

\begin{document}
\maketitle

\begin{abstract}
The Rohingya refugee crisis is one of the biggest humanitarian crises of modern times with more than 600,000 Rohingyas rendered homeless according to the United Nations High Commissioner for Refugees.
While it has received sustained press attention globally, no comprehensive research has been performed on social media pertaining to this large evolving crisis. In this work, we construct a substantial corpus of YouTube video comments (263,482 comments from  113,250 users in 5,153 relevant videos) with an aim to analyze the possible role of AI in helping a marginalized community. Using a novel combination of multiple Active Learning strategies and a novel active sampling strategy based on nearest-neighbors in the comment-embedding space, we construct a classifier that can detect comments defending the Rohingyas among larger numbers of disparaging and neutral ones. We advocate that beyond the burgeoning field of hate-speech detection, automatic detection of \emph{help-speech} can lend voice to the voiceless people and make the internet safer for marginalized communities. 
\end{abstract}


\section{Introduction}

On 25th August, 2017 extreme violence was allegedly perpetrated against the Rohingya community in Rakhine state,  Myanmar~\cite{ohchr}. Since then, according to Human Rights Watch~\cite{humanRights}, more than 671,000 Rohingyas have fled Rakhine state to escape persecution. The Myanmar military's alleged large-scale campaign of ethnic cleansing~\cite{beyrer2017ethnic} has led to one of the fastest-growing refugee crises in the 21st century. However, Myanmar's military and civilian officials have repeatedly denied any claims of atrocities - which are contradicted by extensive evidence~\cite{ohchr} and witness accounts indicating widespread genocide or ethnic cleansing.

Modern history has witnessed multiple instances of mass migration of persecuted communities. Our main goal in this paper is not to argue about highly debated issues like Rohingyas' citizenship rights or make politically contentious claims about the Myanmar government, the alleged oppressor's involvement in this crisis. Rather, our goal is to present the first-of-its-kind case-study of what we call a 21st century problem: migrant crisis in the era of the ubiquitous internet, where the global audience can weigh in on the matter, shape public opinion about the persecuted community through social media comments, clamor for justice for the oppressed, mobilize help to the community in distress, and perhaps side with the alleged oppressor and paint a picture of distrust, fear and threat about a persecuted minority. In online forums, persecuted communities may have little or no voice in discussions centered around them because (i) much of the discussion occurs in a global language (e.g., English) in which they may have limited proficiency, or (ii) they may have minimal access to internet, or most importantly, (iii) survival is possibly the highest priority demanding a significant chunk of their resources. Online activities disparaging refugee communities may result in serious real world consequences; prior research has even identified a close, causal link between online hate speech and offline violence targeting refugees~\cite{muller2018fanning}.

\noindent\textbf{Contributions:} \begin{compactenum}
\item \emph{Domain}: In this paper, via a substantial corpus constructed using comments on YouTube videos (5,153 videos, 263,482 comments posted by 113,250  users) relevant to the Rohingya refugee crisis, we characterize several key aspects of the discourse and show that a medium as powerful as the internet can create an asymmetric discourse where an (allegedly) oppressed minority may have little or no voice to defend themselves from (possibly misinformed) global vitriol. To the best of our knowledge, this is the first AI-focused comprehensive analysis of the Rohingya refugee crisis. In the last decade, besides the Rohingya immigrant crisis, the world has witnessed migrant crises in central America~\cite{bbcCentralAmerica}, Venezuela~\cite{unVenezuela}, Italy~\cite{bbcItalian}, and the long-standing Syrian refugee crisis~\cite{syrianTimeline} resulting in the displacement of millions of people. We believe our work in characterizing key aspects of the Rohingya migrant crisis will open the gates for similar AI research in this humanitarian domain.

\item \emph{Voice for the voiceless}: Existing discourse moderation tools on social media platforms focus on minimizing hate speech through deletion of hostile content and/or flagging belligerent members. We argue that beyond the important field of hate-speech detection, automatic identification of user-generated web content that champions the cause of a minority can be equally critical for making the internet a safer, better, and healthier place. For a balanced and nuanced discussion on issues around
oppressed minorities, lending greater visibility to pro-minority voice (e.g., through pinning or highlighting content)  from a large pool of hostile or ambivalent comments is critical and can be greatly facilitated through automatic methods. To this end, we construct a classifier dubbed the \emph{voice-for-the-voiceless} classifier, that  detects content championing the cause of an oppressed minority, in this case the Rohingyas. 

\item \emph{Machine Learning}: We propose two new active sampling techniques. Our \emph{voice-for-the-voiceless} classifier is constructed using a novel nearest neighbor active-sampling technique in the comment embedding space which effectively uncovers rare positives in a negatively-skewed corpus. In addition, we demonstrate that our proposed technique can be extended to another novel nearest neighbor active-sampling sampling technique in the user embedding space to identify sympathetic users and effectively uncover rare positives.
\end{compactenum}

\section{Concise Overview of the Crisis}

In this section, we present a concise timeline to illustrate the sequence of events that led to this massive humanitarian crisis~\cite{rohingyaTimeline}. As shown in the timeline, arguably, the community experienced a long-standing systemic bias which led to this evolving crisis.

\scalebox{1}{

\begin{tabular}{r |@{\foo} p{0.85\textwidth}}

 1982  & In the 135 national ethnic groups listed  in new  citizenship law, Rohingyas are excluded, effectively rendering them stateless.\\
Nov 13, 2010 & Aung San Suu Kyi, opposition leader and Nobel peace prize winner, is released from house arrest.\\
Jun, 2012 & Religious violence erupts in predominantly Rohingya region Rakhine leaving more than 200 dead and close to 150,000 homeless. In the next three years, more than 112,000 flee to Malaysia by boat.\\
2014 & In Myanmar's first census in three decades, Rohingyas are excluded.   \\    
Nov, 2015 & Suu Kyi's party wins in first election ending military rule. Rohingyas are not allowed to vote let alone contest. \\
Oct 9, 2016 & Rohingya insurgent group Arakan Rohingya Salvation Army (ARSA) claims an attack that killed 9 police officers according  to the state media. A massive military crackdown ensues triggering an exodus of 87,000 Rohingyas to Bangladesh.\\
Aug 25, 2017 & State media claims that 12 police officers were killed by ARSA in a coordinated attack on 20 police posts. A large number of Rohingyas flee to Bangladesh as military responds with (allegedly) burning down villages as a part of what they describe as ``clearance operations''.\\

Oct 23, 2017 & Since Aug 25, 2017, a continuous stream of Rohingya refugees arrive in Bangladesh with the refugee count reaching more than 600,000.\\

\end{tabular}
}


\section{Related Work}

\textbf{Hate-speech detection:} There is a growing body of literature on analyzing and detecting hate-speech in social media such as Facebook~\cite{del2017hate}, Twitter~\cite{davidson2017automated, badjatiya2017deep}, Reddit~\cite{chandrasekharan2017you} and YouTube~\cite{dinakar2012common}. While hate-speech detection and subsequent intervention (by moderating content or flagging users) are extremely helpful in maintaining a positive web environment, these tools are inadequate in this setting. In our problem, a persecuted community is largely absent in an overwhelmingly negative discussion about them, possibly due to language barriers or simply because they lack sufficient access to the web or social media. Detecting comments that advocate their cause is crucial in representing their views. 

\noindent\textbf{Active Learning:} We drew inspiration from several existing lines of Active Learning research for constructing our \emph{voice-for-the-voiceless} classifier~\cite{roy2001, baram2003, donmez2007}. Since sequentially labeling and retraining models may not be practically feasible, following~\cite{yang2013buy}, we adopted a batch Active Learning setting to expand our pool of labeled samples. As we shall demonstrate, a large majority of comments are unfavorable to the Rohingyas, making this classification task one with significant class imbalance. Active Learning with class imbalance is a widely studied research problem (see, e.g., \cite{settles2008analysis, nguyen2004active, donmez2008paired, tomanek2009reducing}). Our proposed sampling strategy leverages recent advances in language modeling to obtain comment embeddings~\cite{joulin2017bag, bojanowski2017enriching} and then mines nearest neighbors in the comment embedding space to alleviate the class imbalance issue. Our proposed solution can thus can be considered a skew-specialized Active Learning approach. However, unlike~\cite{ertekin2009learning}, instead of  constructing a synthetic sample, our method yields samples from the actual pool of unlabeled data. In addition to nearest-neighbor sampling and the more traditional Active Learning sampling strategy, uncertainty sampling, we considered minority-class certainty sampling since it was found to be useful in reducing class imbalance in short document classification tasks ~\cite{attenberg2010unified, sindhwani2009uncertainty, khudabukhsh2015building}.

\noindent\textbf{Research on migrant crisis:}  Extensive research on migrant crises including the Rohingya refugee crisis has been performed from a social science perspective~\cite{xChangeOrg, bhatia2018rohingya, milton2017trapped}. In what follows, we focus on relevant literature with an AI emphasis. Large-scale social media analysis of the Syrian refugee crisis to explore social and communicative networks in Twitter has been performed in~\cite{lynch2014syria}. Using a small set of curated Twitter accounts, community detection has been analyzed in~\cite{o2014online}. 
In terms of domain, our work is most similar to~\cite{chowdhury2018sentiment} where a classifier to label comments favorable to the Rohingyas' resettlement in Bangladesh was constructed. Our work is different in terms of scale, focus and analysis in the following ways. First, we consider a substantially larger corpus of 263,482 comments on videos retrieved using high-frequency search queries from 19 different countries (listed in Table~\ref{tab:countries}), whereas~\cite{chowdhury2018sentiment} focuses on 5,000 Bengali tweets generated by Bangladeshis. Presence of multiple countries adds to our linguistic challenges as expression of intent may become more diverse. Second, we provide a comprehensive analysis of the corpus employing topic modeling, user-level analysis to demonstrate under-representation of Rohingya sympathizers in the global discussion, and overall sentiment analysis of the corpus using domain-specific sentiment lexicons. Third, our \emph{voice-for-the-voiceless} classifier is more nuanced than merely the sentiment towards settlement in one particular country. Finally, we address a critical challenge of mining positive examples in a rare-class learning problem with a novel approach of nearest-neighbor sampling. 

\begin{figure}[t]

\centering
\includegraphics[trim={0 0 0 0},clip, width = 0.6 \textwidth]{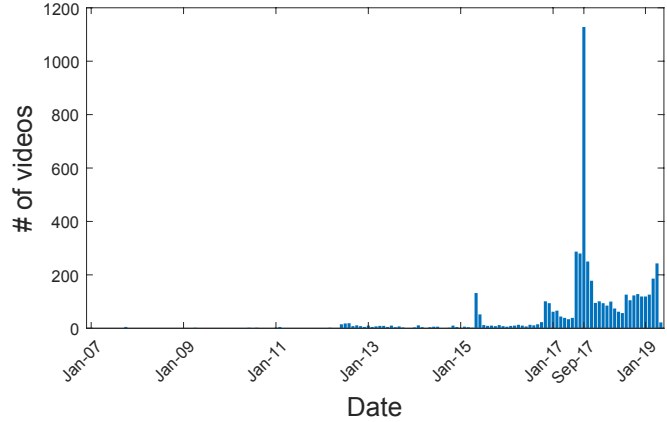}
\caption{Temporal distribution of videos in $\mathcal{V}$.}
\label{fig:videoTemporal}
\end{figure}

\section{Data Collection}


Our data collection process consists of the following steps: 
\begin{compactenum}
    \item We construct a query set, $\mathcal{Q}$ (116 unique queries), by including related queries from Google Trends\footnote{\url{https://trends.google.com/trends/?geo=US}} for the query [\texttt{Rohingya}] from countries listed in Table~\ref{tab:countries}. 
   
\item For each query in $\mathcal{Q}$, we obtained the top 200 YouTube video search results. In addition to these, we performed a targeted crawl focusing on the three months time-duration (July 1, 2017 through September 30, 2017) when the crisis reached its peak. For a given month, for each query in  $\mathcal{Q}$, we obtained the top 50 YouTube video search results. Our final consolidated video data set, $\mathcal{V}$, consists of 5,153 unique videos.

\item  We used the publicly available YouTube API to crawl the comments for the videos obtaining 263,482 comments posted by 113,250  unique users. 


\item Since $\mathcal{Q}$ contains queries from multiple countries where English is not the native tongue, we expected the comment corpus to be a mixed bag of different languages with English being the predominant one. Hence, we required an automated method to identify the English comments. We used an unsupervised language filtering approach~\cite{IndPak}. The size of our English-filtered corpus, denoted as $\mathcal{C}$, is 138,978 comments (i.e., more than 50\% of the corpus was written in English). We identified German, Hindi, Bengali, Malay, Urdu, French, and Arabic in our non-English corpus indicating a  global presence in the discussion.  

\end{compactenum}

\begin{table}[htb]
{
\begin{center}
     \begin{tabular}{|p{5cm} | p{5cm} | }
    \hline
    Countries considered & Reason for inclusion   \\
    \hline
   Myanmar & Origin country of crisis \\
    \hline
    Malaysia, Indonesia and Bangladesh & Countries that offered most help \\
    \hline
    Bangladesh,  China,  Laos, India and Thailand  & Border sharing countries \\
   
    \hline
     Germany, Australia, Austria, Canada, Sweden, Norway, Switzerland, Finland, Ireland  
     and Spain & Granted asylum to Syrian refugees \\
    \hline
    USA & Received 10\% of all asylum applications in the OECD countries in 1998-2007~\cite{haddal2009refugee}.\\ 
    \hline

    \end{tabular}
    
\end{center}
\caption{List of countries.}
\label{tab:countries}}
\end{table}

\noindent\textbf{Video and comment statistics:} We present views, comments and like-related statistics of our videos in Table~\ref{tab:videoStat}. The total number of views of all videos exceeded 15 million. As shown in Figure~\ref{fig:videoTemporal}, a substantial chunk of the obtained videos come from the month of September, 2017, when the crisis reached its peak. Hence, we believe that our video corpus captured most of the relevant coverage on this issue. 

\begin{table}[htb]
{
\begin{center}
     \begin{tabular}{|l | c | }
     \hline 
      & Mean $\pm$ Standard deviation\\
    \hline
    Views     &   29,724 $\pm$ 132,869 \\
    \hline
    Likes &  452 $\pm$ 2100\\
    \hline 
    Dislikes & 36 $\pm$ 145 \\
    \hline
    Comments & 51 $\pm$ 238 \\
    \hline

    \end{tabular}
    
\end{center}
\caption{Statistics related to videos in $\mathcal{V}$.}
\label{tab:videoStat}}
\end{table}

\begin{figure}[htb]

\centering
\subfigure[Rohingyas are]{%
\includegraphics[frame,trim={8mm 15mm 8mm 15mm},clip, width = 0.40 \textwidth]{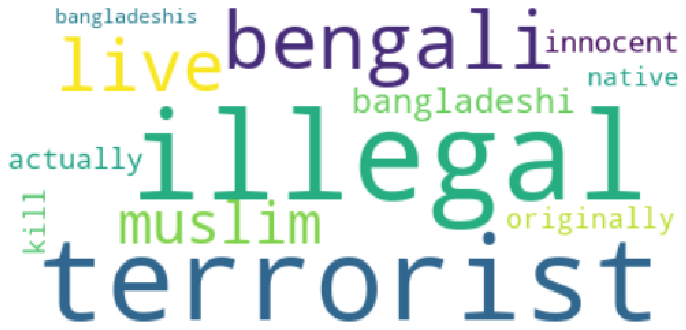}
\label{fig:theyAre}}
\subfigure[Rohingyas are not]{%
\includegraphics[frame,trim={8mm 15mm 8mm 15mm}, clip, width = 0.40 \textwidth]{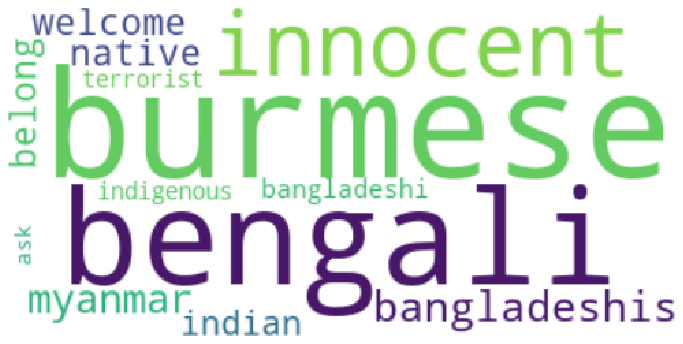}
\label{fig:theyAreNot}}
\caption{{A word cloud visualization of Rohingyas' perception. Among 992,841 unique bigrams and 2,465,453  unique trigrams, in terms of frequency, \texttt{[Rohingyas are]} and \texttt{[Rohingyas are not]} rank 188$^{th}$ and 358$^{th}$, respectively.}}
\label{fig:perception}
\end{figure}

\noindent\textbf{General perception:} We first analyze phrases in the comments
that match a set of high-frequency text templates. We  focus on: \texttt{[Rohingya are]} (or \texttt{[Rohingyas are]}) and the negated variants (\texttt{[Rohingya are not]} and \texttt{[Rohingyas are not]}). The tokens that follow these templates are visualized in Figure~\ref{fig:perception} demonstrating that the prevalent perception of Rohingyas was largely negative. The templates were chosen by observing the top phrases
that occurred in the corpus. A substantial majority often equated them with terrorists and only a tiny fraction of comments stated they were innocent. When the negated versions were used (Figure~\ref{fig:theyAreNot}) however, we noticed that \texttt{innocent} became a high-frequency term. Hence, their innocence is questioned by general commenters. Additionally,  Figure~\ref{fig:theyAreNot} indicates a debate around
the ethnicity of the Rohingyas - several commenters stated that they are neither Burmese nor Bengali,
which leads to our next analysis.

\begin{figure}[htb]
\centering
\includegraphics[trim={8mm 15mm 8mm 15mm},clip,frame, width = 0.40 \textwidth]{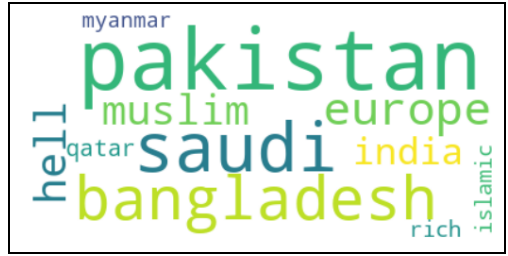}
\label{fig:sendThemTo}
\caption{{Resettlement debate. Among 2,465,453 unique trigrams, in terms of frequency, \texttt{[Send them to]} ranks 72$^{nd}$ respectively.}}
\end{figure}

\begin{table}[htb]
  \centering
  \begin{tabular}{|l|c|c|}
    \hline
    Topic 1 (77.5\%) & Topic 2 (16.3\%) & Topic 3 (3.2\%) \\
    \hline
    people     & allah    & child      \\
    muslims    & quran    & rapist     \\
    rohingya    & god      & aisha      \\
    myanmar      & muhammad & puberty    \\
    india    & islam    & mentally   \\
    bangladesh & prophet  & muhammedans \\
    \hline
    \end{tabular}
    \vspace{0.2cm}
    \caption{{Most relevant tokens for three major topics discovered in the Rohingya corpus using~\cite{blei2003latent}}.}
  \label{tab:rohingyalda}
\end{table}

\noindent\textbf{The resettlement debate:} We focused on the template [\texttt{send them to}] to analyze public perception
of where they should be resettled and which community or country is responsible for providing assistance and protection. Figure 3 shows
that the issue of origin and resettlement is a highly discussed issue in the corpus. Apart from rich Muslim countries, and obvious choices like neighboring countries India and Bangladesh, 
we were alarmed to notice that [\texttt{send them to hell}] was also a frequently occurring 4-gram in the corpus; among 3,215,489 unique 4-grams, its percentile rank is 99.75.

\noindent\textbf{Aspects of the discussion:}
We next employ topic modeling to obtain a fuller picture of the discussion in the corpus.
For our topic model analysis, we first filtered comments with 15 tokens or more and then ran the LDA algorithm \cite{blei2003latent} obtaining best results for topic count of 4 (the 4th topic contained code-mixed incoherent tokens). We discover three main themes: (i) a geopolitical
discussion centered around India, Bangladesh, and Myanmar - the three geographically significant countries in the area, (ii) a
religion-centric discussion with an appeal for help, and (iii) an anti-Islamic cluster primarily consisting of religion-themed slurs. Table \ref{tab:rohingyalda} contains a list of the most relevant terms per topic. The relevance score is from \cite{sievert2014ldavis}.

\noindent \textbf{User level analysis}: We were curious to examine: \emph{is it possible to estimate how many Rohingyas engage with videos where global users post highly negative comments about them?} It is not possible to unambiguously identify if a YouTube user is Rohingya or not. 
However, we identified several Rohingya-focused YouTube channels many of which use the Rohingya language (this language has 1.8 million native speakers as compared to 380 million native speakers of English). Users who predominantly commented on videos hosted by such channels could possibly be Rohingya or Rohingya-sympathizers. Consequently, we divided our set of videos into two mutually exclusive sets:  videos that are hosted by Rohingya-focused channels (e.g., Voice of Rohingya) denoted as $\mathcal{V}^{\emph{Roh}}$ (1,727 videos from 123 channels), the other, denoted as $\mathcal{V}^{\emph{other}}$ is the complement of $\mathcal{V}^{\emph{Roh}}$. Videos belonging to $\mathcal{V}^{\emph{other}}$ (3,426 videos from 1,244 channels) are primarily hosted by News channels (e.g.. BBC, Al Jazeera, CNN) and a few individual contributors. 

Of the 113,250 users, 11,326  and 104,973 users commented on $\mathcal{V}^{\emph{Roh}}$ and $\mathcal{V}^{\emph{other}}$, respectively. The overlap between the two sets was 3,049  users (Jaccard similarity index 0.0269). Due to disparate size of the two sets, we admit that instead of looking at Jaccard similarity, a more interesting follow up research question could be \emph{who posts more negative comments? Is it the users who are frequent visitors of} $\mathcal{V}^{\emph{Roh}}$ \emph{but occasionally visit} $\mathcal{V}^{\emph{other}}$\emph{? Or the other way around?} We focus on the 3,049  users who have commented on at least one video belonging to $\mathcal{V}^{\emph{Roh}}$ and one video belonging to $\mathcal{V}^{\emph{other}}$ and define two mutually exclusive user sets: $\mathcal{U}^{\emph{Roh} \rightarrow \emph{other}}$ (users with more than 80\% of comments
posted on videos in $\mathcal{V}^{\emph{Roh}}$) and
$\mathcal{U}^{\emph{other} \rightarrow \emph{Roh}}$ (users with more than 80\% of comments posted on videos in $\mathcal{V}^{\emph{other}}$).

\begin{table}[htb]

{
\begin{center}
     \begin{tabular}{| p{5cm}  | p{5cm} |}
    \hline
    Positive  & Negative\\
     \hline
     peace, great, peaceful, accept, kind, thank, love, care,  humanity, innocent  & terrorist, genocide, war, hate, bad, violence, rape, illegal, evil, shame  \\
    \hline
    \end{tabular}
\end{center}
\caption{Manually selected seed words for \texttt{SENTPROP}.}
\label{tab:seedWords}}
\end{table}

We next obtained English comments made by these two user sets $\mathcal{U}^{\emph{Roh} \rightarrow \emph{other}}$ (denoted by $\mathcal{C}^{\emph{Roh} \rightarrow \emph{other}}$) and $\mathcal{U}^{\emph{other} \rightarrow \emph{Roh}}$ (denoted by $\mathcal{C}^{\emph{other} \rightarrow \emph{Roh}}$ ). Lexicon-based sentiment analysis is an established tool for computing
sentiment scores~\cite{o2010tweets}. In this scheme, tokens are assigned scores and individual documents' (comments in our case) scores are obtained by combining the constituent token scores. For effective sentiment analysis, obtaining a domain-specific lexicon is crucial~\cite{velikovich2010viability}.
We considered two existing lexicons induced on popular sub-reddits~\cite{hamilton2016inducing} (\texttt{politics} and \texttt{India} sub-reddits) and a new custom lexicon induced on our corpus
using 100-dimensional \texttt{FastText} embeddings~\cite{joulin2017bag} and a lexicon inducing algorithm (\texttt{SENTPROP}) ~\cite{hamilton2016inducing}. \texttt{SENTPROP} requires a set of positive and negative seed words (listed in Table~\ref{tab:seedWords}). Our test for positive or negative adds the individual token scores and if
the cumulative comment score is greater than 3 (or less than -3), the comment is considered positive (or negative). 

As shown in Table~\ref{tab:sentiment}, across all three lexicons, we found that $\mathcal{U}^{\emph{Roh} \rightarrow \emph{other}}$ posted substantially fewer negative comments than positive comments in comparison to $\mathcal{U}^{\emph{other} \rightarrow \emph{Roh}}$ where
the ratio of positive to negative comments was reversed.  Human evaluation on
a random sample of 200 comments revealed that a larger share of negative comments posted by $\mathcal{U}^{\emph{other} \rightarrow \emph{Roh}}$ were disparaging to Rohingyas, and the small fraction of negative 200 comments posted by $\mathcal{U}^{\emph{Roh} \rightarrow \emph{other}}$ were mostly against the Myanmar government's (alleged) atrocities. 

\begin{table}[htb]

{
\begin{center}
     \begin{tabular}{|l | c  | c | c |}
    \hline
     & Politics    & India & Induced on  \\
       & sub-reddit & sub-reddit  & $\mathcal{C}$ \\
    \hline
   & pos = \textbf{30.40\%}  & pos = \textbf{25.20\%} & pos = \textbf{48.57\%} \\
   $\mathcal{C}^{\emph{Roh} \rightarrow \emph{other}}$          & neg = 2.38\%  & neg = 1.85\% & neg = 2.88\%\\
    \hline
    & pos = 10.01\%,   & pos = 12.45\% & pos = 18.55\%\\
   $\mathcal{C}^{\emph{other} \rightarrow \emph{Roh}}$              &  neg = 32.92\%  & neg = 32.76\% & neg = 35.30\%\\ 
    \hline

    \end{tabular}

\end{center}
\caption{Percentage of positive and negative comments using lexicons presented in ~\cite{hamilton2016inducing} and a lexicon induced
on $\mathcal{C}$.}
\label{tab:sentiment}}
\end{table}

\section{Voice for the Voiceless Classifier}

We start with pointing out a subtle but important distinction: \emph{Voice-for-the-voiceless speech is not absence of hate speech}. The goals of a hate-speech classifier and our voice-for-the-voiceless classifier are different and complement each other. Identifying hateful content for possible moderation certainly has a positive role in making the internet a safer place for a vulnerable community. However,  surfacing comments marked as  \emph{\textbf{not} hate speech} does not necessarily lend a voice to the voiceless. For instance, say a user from India respectfully states that India is an over-populated country and does not have enough resources for Rohingyas. This is clearly not hate-speech against the Rohingyas, but it is also not voicing the concerns of the voiceless (the Rohingyas). 

We next present a definition of \emph{voice-for-the-voiceless} speech  and provide examples picked from the corpus or written by us (italicized) to illustrate the point. Understandably, the italicized comments succinctly express a given condition in correct English while examples from the corpus might contain grammatical errors.

\noindent\textbf{Definition 1:} A comment is marked as  \emph{voice-for-the-voiceless} speech, if 
the comment
\begin{compactenum}
\item actively seeks to help one or more persons belonging to the (allegedly) oppressed minority (e.g., \texttt{[how can we help the Rohingyas]})  
\item urges other people or organizations (such as the UN) to help the (allegedly) oppressed minorities (e.g., \emph{UN should help Rohingyas})
\item urges other people to come forward and assist or take a humane stance (e.g.,  \texttt{[value the humanity thy r migrating for lives not for luxury]})
\item advocates for the (allegedly) oppressed community's rights (e.g., \emph{Rohingyas should get Myanmar citizenship})
\item condemns the atrocities against the (allegedly) oppressed (e.g., \emph{Myanmar government shame on you})  
\item sympathizes with the (allegedly) oppressed community's plight (e.g., \texttt{[This just breaks my heart. I wish I could help. All these people commenting about muslims and hindus should be ashamed.  The bottom line is these are humans being killed, children being killed. It doesnt matter who started it. It needs to stop!]})
\end{compactenum}
    
\noindent and a comment is \textbf{not} \emph{voice-for-the-voiceless} if it

\begin{compactenum}

\item expresses violent intent to a specific entity (including the alleged oppressors) or broad bias against any religious community (e.g., \texttt{[Pakistan please nuke Myanmar bhudists]})
\item calls for aggressive action against the oppressed community (e.g., \texttt{deport them all})
\item demonstrates proverbial whataboutism (e.g., \texttt{[what about Yazidis]})
\item paints a general picture that the community is a threat (e.g.,\texttt{[Rohingyas are terrorists]})
\item shows solidarity with the (alleged) oppressors (e.g., \texttt{[well done Myanmar]})

\end{compactenum}

\noindent\textbf{Active learning with class imbalance:} Typically, in Active Learning, a \emph{seed set} of samples is used to construct a classifier which then samples from the unlabeled pool and seeks labels~\cite{settles2009active}. For better generalizability of the classifier in the wild, it is often critical that the training set is (i) balanced, i.e., contains sufficient number of examples from both classes (ii) diverse, i.e. captures a wide variety of data points we may encounter in the wild. 

We faced the following two research challenges:
\begin{compactitem}
    \item How to obtain a sufficient number of positive comments in a corpus largely disparaging to the Rohingyas?
    \item How to cover a wide range of aspects of positive (and also negative) comments  in our training set so that the classifier performs well in the wild?
\end{compactitem}

\noindent\textbf{Active Learning meets document embeddings:} Note that, key phrases (e.g., \texttt{send them to}, \texttt{deport them all}, \texttt{breaks my heart)} may express user intent. However, in a corpus largely filled with negative comments and with a high variance in English proficiency among the contributors, simple mining techniques using exact phrase-level match may not yield sufficient number of positives. In the extreme case, the phrase we are looking for, may not even yield a single exact match. Moreover, the matched comments run the risk of being highly similar to each other and hence may not capture the entire space of varied expressions. Using semantic embeddings to find  comments similar to an example positive phrase (or negative phrase) may be effective; however, with a smaller corpus, semantic embeddings may be more prone to inaccuracy (alleviated with a human-in-the-loop in the Active Learning setting). In this work, we meld recent advances in sentence embeddings with Active Learning and propose a novel Active Sampling technique to augment our seed set. The model described in \cite{pgj2017unsup} is used to obtain a real-valued vector for each comment
in the corpus and used to retrieve a comment's nearest neighbors in this embedding space. In conjunction with random sampling, this technique helps us discover a broader, more diverse set of positive examples and
helps us combat extreme class imbalance.

\subsection{Our Active Learning Approach}

\begin{figure}[htb]
\centering
\includegraphics[trim={0 0 0 0},clip, scale = 0.38]{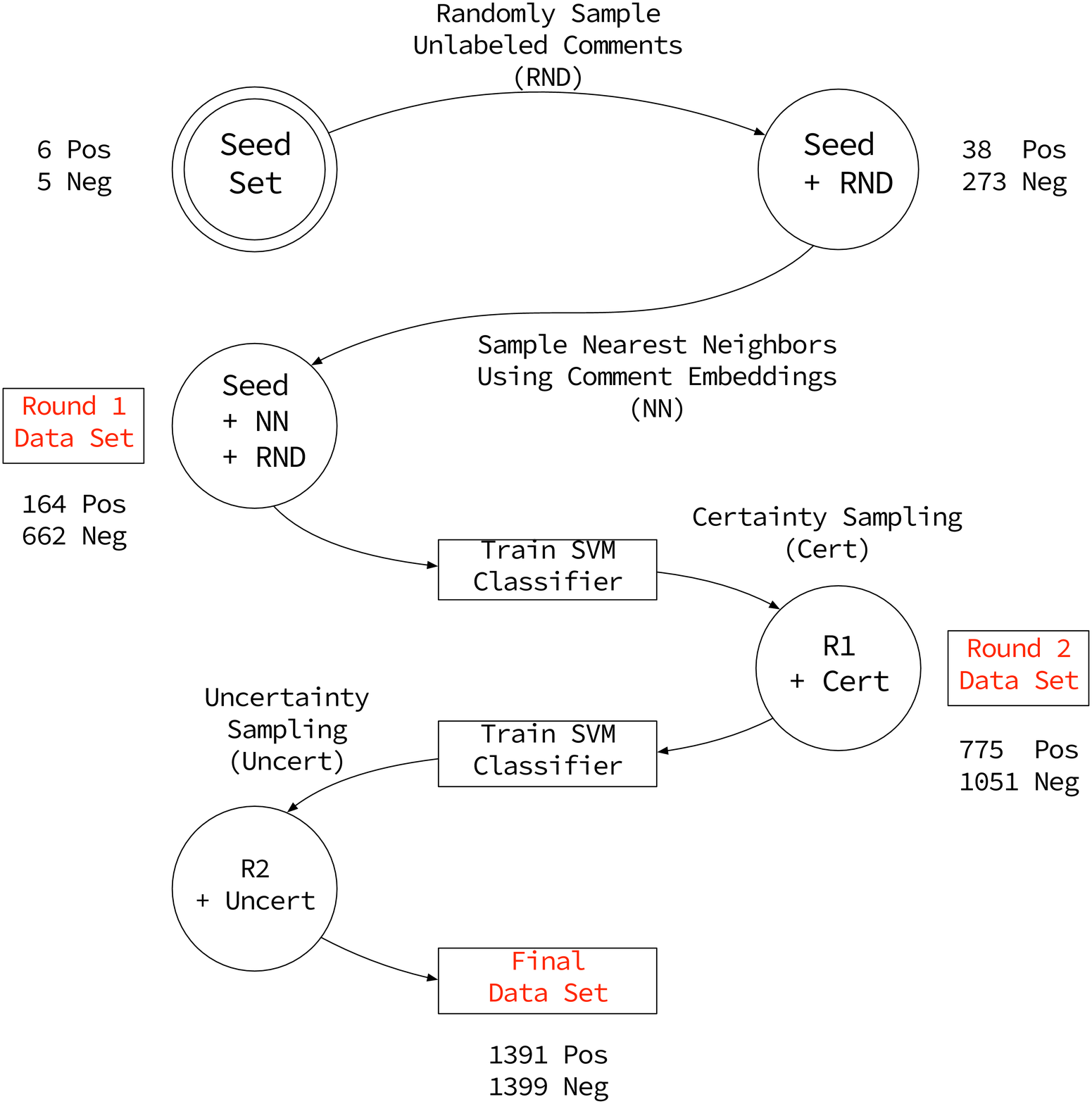}
\caption{System diagram.}
\label{fig:ALSystemFlow}
\end{figure}

As illustrated in Figure~\ref{fig:ALSystemFlow}, our approach consists of the following steps:
\begin{compactenum}
    \item Construct a seed set of positive and negative comments.
    \item Expand the seed set by randomly sampling comments from the unseen corpus.
    \item Obtain real valued embeddings for the comments, find the nearest neighbors of the seed set and include them in the corpus (new technique presented in this paper).
    \item Further expand using minority-class certainty sampling.
    \item Perform final expansion using uncertainty sampling.
\end{compactenum}

\hide{
\begin{table}[htb]
{
\smaller
\begin{center}
     \begin{tabular}{|l | c |}
    \hline
    Technique & Fraction of positives  \\
    \hline
   Random Sampling & 10.66\%   \\
    \hline
   NN in the comment embedding space & \textbf{34.58}\%    \\
    \hline
    \end{tabular}

\end{center}
\caption{Sampling performance.}
\label{tab:NNSamplingPerf}}
\end{table}
}

\noindent\textbf{Seed set:} Our seed set (6 positives, 5 negatives) consists of the same set of examples presented in Definition 1.  

\noindent\textbf{Random sampling:} In order to have better coverage, we randomly sampled 300 comments and labeled them. We obtained 32 positives and 268 negatives, i.e., 10.66\% positives. For evaluating our sampling strategies this acts as the baseline.  All rounds of manual labeling were performed by two annotators proficient in English. The annotators were presented with the definition and example seed set. They were first asked to label independently, and then allowed to discuss and resolve the disagreed labels. We obtained strong agreement in every round (lowest Cohen's $\kappa$ coefficient across all rounds was 0.8766 indicating strong inter-rater agreement).

\noindent\textbf{Nearest-neighbor sampling (NN sampling):} We use a well-known document embedding model from \cite{pgj2017unsup} to obtain a real-valued embedding for each comment in our corpus. Starting from the seed set, we obtained the seed embeddings and then obtained the comments
from the unlabeled corpus whose embeddings were closest to the seed embeddings (i.e. the nearest neighbors). following ~\cite{demszky2019analyzing, pgj2017unsup} cosine distance
was used as the distance metric.

\begin{table}[htb]
{
\begin{center}
     \begin{tabular}{|p{7.1cm}|}
    \hline
    can someone tell me where i can \textcolor{blue}{help} charity to them  \\
    \hline
   all the countries should take a stand for these people and force mayanmar government to \textcolor{blue}{accept them}   \\
    \hline
   No country is too small to \textcolor{blue}{take on refugees} and camp them for period of time until the problem is solved by the world leaders making every problem a political issue is just creating dangerous matters for the poor public in some countries animals are cared for more than the humans    \\
   \hline 
   \textcolor{blue}{sanction myanmar} till they understand international law and \textcolor{blue}{give up ethnic cleansing}\\
    \hline
    \end{tabular}

\end{center}
\caption{{Random sample of positive comments obtained using the nearest-neighbor sampling.}}
\label{tab:samples}}
\end{table}

\noindent\textbf{Advantages of our technique:} First, it allows flexibility while specifying an example comment. Without sufficient knowledge of the corpus, it may be difficult to uncover a rare positive satisfying a particular aspect of the target concept. Some of the examples in our seed set (italicized) did not have an exact match in the actual corpus yet the semantic similarity technique uncovered similar rare positives. Second, our embedding method \cite{pgj2017unsup} employs sub-word information and thus is robust to spelling variations or outright erroneous spellings (e.g., Buddhists was incorrectly spelled as bhudists). In addition, it can handle both short and long comments. In Table~\ref{tab:samples}, we list a few positives we obtained through our technique to highlight its effectiveness. Out of the 300 nearest neighbors obtained from 6 positive seed comments, we obtained 101 unique positives (33.67\%) from 292 unique comments. The large number of unique comments indicates that our technique found a diverse set of samples. We obtained more than 3x the number of positives than discovered by random sampling (10.66\%). Our method also uncovered a diverse set of negative comments about the Rohingyas; a representative sample is listed in Table~\ref{tab:samplesNegative} to emphasize why we believed the community required protection from online attacks. 

We next trained a Support Vector Machine (SVM) classifier on our consolidated labeled data set with 164 unique positives and 662 unique negatives (we also included the randomly sampled instances) with token unigram, bigram and trigram features.

\begin{table}[t]
{
\begin{center}
     \begin{tabular}{|p{7cm}|}
    \hline
    rohingyas are very \textcolor{red}{strong in breeding} \\
    \hline
   rohingya muslims \textcolor{red}{are terrorists} they have been killing buddhist from 1947 onwards \textcolor{red}{they deserve} whatever they are getting   \\
    \hline
   kick them all out \textcolor{red}{fuc ing like swines} and \textcolor{red}{changing our demographics}     \\
   \hline 
   just \textcolor{red}{kill them all} soon because they are \textcolor{red}{terrorists bastards}\\
    \hline
    \end{tabular}

\end{center}
\caption{{Random sample of negative comments obtained using the nearest-neighbor sampling.}}
\label{tab:samplesNegative}}
\end{table}

\begin{table}[t]
{
\begin{center}
     \begin{tabular}{|l | c | c | c | c |}
    \hline
    Performance & Seed set + random sampling +    & + Certainty & + Uncertainty  \\
    measure  & NN in the embedding space & sampling & sampling\\
    \hline
   Precision & 67.17 $\pm$ 9.90\%  & 71.27 $\pm$ 5.23\% &  73.65 $\pm$ 3.45\%\\
    \hline
    Recall & 32.35 $\pm$ 7.65\%  & 72.52 $\pm$ 4.23\%  &  79.39 $\pm$ 3.72\%\\
    \hline
     Accuracy & 82.04 $\pm$ 2.34\%  & 75.95 $\pm$ 3.10\% &  75.38 $\pm$ 2.76\%\\
    \hline
     F1 score & 43.02 $\pm$ 7.90\%  &  71.75 $\pm$ 4.32\% & 76.34 $\pm$ 2.77\%\\
    \hline
    AUC & 83.61 $\pm$ 2.88\% &  83.64 $\pm$ 2.84\% &  83.67 $\pm$ 2.61\%\\
    \hline
    
    \end{tabular}
    
\end{center}
\caption{\emph{Voice-for-the-voiceless} classifier performance.}
\label{tab:classifier}}
\end{table}

\begin{table}[t]
{
\begin{center}
     \begin{tabular}{|l | c | c |}
    \hline
    Performance & SVM (n gram)    & SVM (n gram + embedding) \\
    measure & & \\
    \hline
    Precision  &73.65 $\pm$ 3.45\% & \textbf{76.49} $\pm$ 3.51 \\
    \hline 
    Recall &79.39 $\pm$ 3.72\% & \textbf{80.30} $\pm$ 3.73 \\
    Accuracy &75.38 $\pm$ 2.76\%& \textbf{77.71} $\pm$ 2.56 \\
    \hline 
    F1 score &76.34 $\pm$ 2.77\%& \textbf{78.28} $\pm$ 2.71 \\
    \hline
    AUC &83.67 $\pm$ 2.61\%& \textbf{85.91} $\pm$ 2.32 \\
    \hline

    \end{tabular}
    
\end{center}
\caption{\small{Model improvement.}}
\label{tab:modelImprovement}}
\end{table}

\noindent\textbf{Certainty sampling.} While our \emph{NN sampling} technique effectively uncovered a considerable number of rare positives, the class imbalance was still present with positives merely constituting 19.85\% of the labeled data set. We bridged this gap through employing \emph{certainty sampling}, a sampling technique first proposed in~\cite{sindhwani2009uncertainty, attenberg2010unified}. In batch certainty sampling, we pick $k$ (set to 1000) unlabeled samples with highest predicted probability for the minority class. In this step, we closed the gap between the number of positives and negatives as we obtained 611 unique positives and 389 unique negatives. We re-train our classifier with our consolidated data set of 775 positives and 1,051 negatives.

\noindent \textbf{Uncertainty sampling.} Finally, we used uncertainty sampling to add 1000 more samples where the predicted class probability was close to 0.5. Our final data set consists of 2790 comments with roughly equal numbers of positives and negatives (1,391 positives and 1,399 negatives). Hence, via (i) active learning, (ii) combining multiple existing sampling techniques and (iii) our proposed nearest-neighbor sampling, we succeeded in addressing class imbalance.

\begin{table}[htb]
{
\begin{center}
     \begin{tabular}{|p{13cm}|}
    \hline
    \textcolor{red}{keep helping these poor innocent people the myanmar government is really kind of like \textbf{animals} not like human beings so thats why they genocide innocent people in myanmar}  \\
    \hline
   i am from nepal where buddha was born i have seen buddhist who is so kindful n helpful but i never seen buddhist who murder poor n innocent people i really fell so shameful that they are killing innocent poor people children and old age people they are torturing kid and womens for god shake please stop this violence   \\
    \hline
   \textcolor{blue}{thank you so for news today and vi want full human rights in arakan myanmar and stop nvc card and ples vi want myanmar army goverment to the icc kireminal courd justice and vi want full setizenthip in arakan myanmar vi no bangali vi setizenthip in arakan myanmar and myanmar army reped womens rohingya and etnik kilingsing of rohingya and genocide of rohingya and ples vi want hlep from un konsiel and from human rights wohc ples hlep stop genocide of rohingya and humanty in myanmar and thank you so lot god bles you all}\\
   \hline 
   its genocide ethnic cleansing brutality reach the level of where words cannot describe its inhuman government of myanmar monk are killing babies and womens into pieces sushi should be punished by court of law\\
    \hline
    \end{tabular}

\end{center}
\caption{{Performance in the wild.}}
\label{tab:wild}}
\vspace{-.1in}
\end{table}

\noindent\textbf{Classifier performance.} We used a 90/10 train/test split; we trained an SVM classifier~\cite{tong2001support} with token n-grams as features (with $n$ up to 3) and evaluated the performance on the test set. Since performance can be sensitive to individual test/train splits, we repeated the experiment 100 times on 100 randomly chosen test-train splits. Our intermediate classifiers and final classifier's performance is summarized in Table~\ref{tab:classifier}. In a class-imbalanced problem, simply predicting the majority class can yield a high accuracy, F1 score is the more meaningful measure. After each round of labeled data acquisition, we noticed a steady rise in the F1 score with a final performance of 76.50 $\pm$ 2.85\%. We obtain further improvement by adding comment embeddings as features as shown in Table~\ref{tab:modelImprovement}.

\begin{figure}[htb]
\centering
\includegraphics[trim={0 0 0 0},clip, width=4.1in, height = 2.3in]{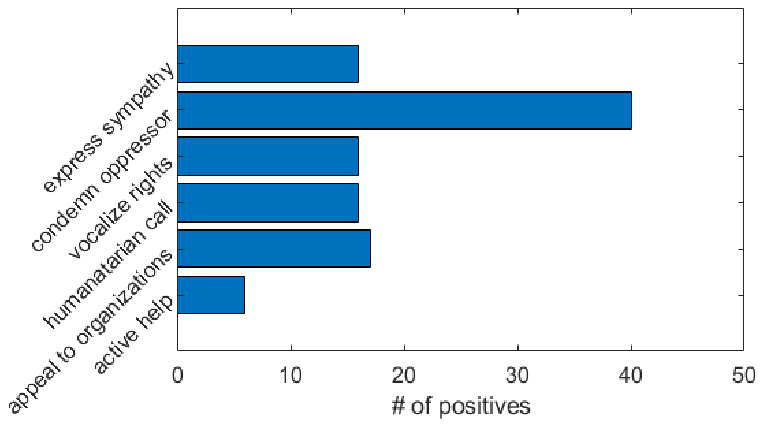}
\caption{\small{Breakdown of positive comments found in the wild. A single comment can satisfy multiple criteria.}}
\label{fig:distribution}
\vspace{-0.2cm}
\end{figure}

\hide{Since our prediction task is novel, we do not have any existing baselines to compare against. However, we looked up previous literature on hate-speech detection~\cite{djuric2015hate, davidson2017automated}. Compared to~\cite{djuric2015hate}, where the reported AUC (Area Under the Curve) in hate-speech detection of  user comments collected on Yahoo Finance website was 80.07\%, we obtained slightly better performance with an AUC of 83.61\% on a significantly more challenging corpus and a more nuanced target concept.  \cite{davidson2017automated} presented a fine-grained identification of hate-speech with an aim to distinguish hate-speech from plain offensive language. Compared to the precision (44\%) and recall (61\%) reported for the hate class in~\cite{davidson2017automated}, we achieved a final precision and recall of 73.86\% and 79.47\%, respectively.}

\noindent\textbf{Performance in the wild:} Our goal is to identify comments supporting a persecuted minority in the wild. We ran our classifier on the unlabeled corpus (i.e., on comments neither belonging to the train or test set) and  conducted a human evaluation of the top 100 comments predicted as \emph{voice-for-the-voiceless} ranked by confidence. Of the 100 comments, 88$\%$ were annotated as positives which indicates the classifier would substantially reduce manual effort (random sampling only found 10.66$\%$ positives) to find supportive comments for Rohingyas in the wild. In Figure~\ref{fig:distribution}, we present the breakdown of the positives into six broad categories as presented in our definition (active help, appeal to organizations, humanitarian call, vocalize rights, condemn oppressor, express sympathy). We found that our classifier found comments belonging to all broad categories from the wild. In Table~\ref{tab:wild}, we highlight few randomly sampled comments to illustrate two points. First, we draw attention to the comment highlighted in blue. We suspect that this comment is written by a Rohingya YouTube user. Broken sentences, grammatical disfluency and a large number of spelling errors indicate how the language barrier may make it difficult for a marginalized community to voice their opinion. Our classifier correctly labeled this comment with high confidence, indicating our approach holds promise in surfacing minority voices. However, since the minority is experiencing (alleged) persecution, it is reasonable to observe measured expression of rational negativity while condemning the (alleged) oppressor. The comment marked in red opens up an interesting philosophical question: \textbf{\emph{where should we draw the line?}} For instance, one particular comment supporting the Rohingyas found in the wild used a gendered insult to refer to the Prime Minister of Myanmar which our annotators marked as negative. Hence, we conclude this analysis by saying that our classifier holds promise to substantially lessen the burden of moderators to automatically find content supporting a minority, however it may require some further supervision and human judgement to ensure fairness. This research question merits deeper exploration.

\subsection{\emph{Voice-for-the-voiceless} community}

We conclude our paper with a small exploratory study on the possibility of finding rare positives through user embeddings. For a given user, we constructed the corresponding user
embedding  using the embeddings of all comments posted by the user, normalizing them and finally averaging
these normalized embeddings. Our user-focused nearest-neighbor sampling consists of the following steps:\\ (1) Obtain top $k$ positive comments (ranked by predicted class probability) predicted by the \emph{voice-for-the-voiceless} classifier. (2) Next, identify the set of unique users, $\mathcal{U_{\emph{top}}}$, who posted these comments. (3) Next, for each user in $\mathcal{U_{\emph{top}}}$, obtain $m$ nearest neighbors in the user embedding space. (4) Finally, sample comments from the nearest neighbors.


We set both $k$ and $m$ to 10. We obtained 9 unique users who posted the top 10 comments. Of the 90 nearest neighbors, 88 were unique indicating our user-focused nearest neighbor sampling was able to uncover a diverse set of users. We next randomly sampled 300 comments posted by the nearest neighbors. Our hypothesis was if our user embedding-based sampling indeed identifies a set of positive users, the sampled comments will have more positives than the baseline (random sampling fetched 10.66\% positives). Our annotators identified 105 positives (35\%). Hence, our user-focused sampling performed 3x better than the baseline. Hence, both Active Sampling strategies proposed in this paper substantially outperformed the baseline in finding rare positives supporting a persecuted minority. We conclude with the hope that this work will motivate further AI research in this important humanitarian domain.

\bibliographystyle{unsrt}

\begin{thebibliography}{10}

\bibitem{ohchr}
Nathan Thompson.
\newblock Myanmar: Un fact-finding mission releases its full account of massive
  violations by military in rakhine, kachin and shan statesa, 2018.
\newblock [Online; accessed 12-May-2019].

\bibitem{humanRights}
Human~Rights Watch.
\newblock Rohingya crisis.
\newblock [Online; accessed 12-May-2019].

\bibitem{beyrer2017ethnic}
Chris Beyrer and Adeeba Kamarulzaman.
\newblock Ethnic cleansing in myanmar: the rohingya crisis and human rights.
\newblock {\em The Lancet}, 390(10102):1570--1573, 2017.

\bibitem{muller2018fanning}
Karsten M{\"u}ller and Carlo Schwarz.
\newblock Fanning the flames of hate: Social media and hate crime.
\newblock {\em Available at SSRN 3082972}, 2018.

\bibitem{bbcCentralAmerica}
BBC News.
\newblock Us-mexico border official says migrant crisis `at breaking point',
  2019.
\newblock [Online; accessed 12-May-2019].

\bibitem{unVenezuela}
Mark~Leon Goldberg.
\newblock Venezuela is a refugee crisis, 2019.
\newblock [Online; accessed 12-May-2019].

\bibitem{bbcItalian}
BBC News.
\newblock Italy migrant crisis: Government passes tough bill, 2019.
\newblock [Online; accessed 12-May-2019].

\bibitem{syrianTimeline}
Office of~the United Nations High Commissioner~for Refugees.
\newblock Seven years on: Timeline of the syria crisis, 2018.
\newblock [Online; accessed 12-May-2019].

\bibitem{rohingyaTimeline}
Katie Hunt.
\newblock Rohingya crisis: How we got here, 2017.
\newblock [Online; accessed 12-May-2019].

\bibitem{del2017hate}
Fabio Del~Vigna12, Andrea Cimino23, Felice Dell’Orletta, Marinella Petrocchi,
  and Maurizio Tesconi.
\newblock Hate me, hate me not: Hate speech detection on facebook.
\newblock In {\em Proceedings of the First Italian Conference on
  Cybersecurity}, 2017.

\bibitem{davidson2017automated}
Thomas Davidson, Dana Warmsley, Michael Macy, and Ingmar Weber.
\newblock Automated hate speech detection and the problem of offensive
  language.
\newblock In {\em Eleventh International AAAI Conference on Web and Social
  Media}, 2017.

\bibitem{badjatiya2017deep}
Pinkesh Badjatiya, Shashank Gupta, Manish Gupta, and Vasudeva Varma.
\newblock Deep learning for hate speech detection in tweets.
\newblock In {\em Proceedings of the 26th International Conference on World
  Wide Web Companion}, pages 759--760. International World Wide Web Conferences
  Steering Committee, 2017.

\bibitem{chandrasekharan2017you}
Eshwar Chandrasekharan, Umashanthi Pavalanathan, Anirudh Srinivasan, Adam
  Glynn, Jacob Eisenstein, and Eric Gilbert.
\newblock You can't stay here: The efficacy of reddit's 2015 ban examined
  through hate speech.
\newblock {\em Proceedings of the ACM on Human-Computer Interaction},
  1(CSCW):31, 2017.

\bibitem{dinakar2012common}
Karthik Dinakar, Birago Jones, Catherine Havasi, Henry Lieberman, and Rosalind
  Picard.
\newblock Common sense reasoning for detection, prevention, and mitigation of
  cyberbullying.
\newblock {\em ACM Transactions on Interactive Intelligent Systems (TiiS)},
  2(3):18, 2012.

\bibitem{roy2001}
Nicholas Roy and Andrew McCallum.
\newblock Toward optimal active learning through monte carlo estimation of
  error reduction.
\newblock {\em ICML, Williamstown}, pages 441--448, 2001.

\bibitem{baram2003}
Yoram Baram, Ran~El Yaniv, and Kobi Luz.
\newblock Online choice of active learning algorithms.
\newblock {\em Journal of Machine Learning Research}, 5(Mar):255--291, 2004.

\bibitem{donmez2007}
Pinar Donmez, Jaime~G Carbonell, and Paul~N Bennett.
\newblock Dual strategy active learning.
\newblock In {\em European Conference on Machine Learning}, pages 116--127.
  Springer, 2007.

\bibitem{yang2013buy}
Liu Yang and Jaime Carbonell.
\newblock Buy-in-bulk active learning.
\newblock In {\em Advances in neural information processing systems}, pages
  2229--2237, 2013.

\bibitem{settles2008analysis}
Burr Settles and Mark Craven.
\newblock An analysis of active learning strategies for sequence labeling
  tasks.
\newblock In {\em Proceedings of EMNLP}, pages 1070--1079, 2008.

\bibitem{nguyen2004active}
Hieu~T Nguyen and Arnold Smeulders.
\newblock Active learning using pre-clustering.
\newblock In {\em Proceedings of the twenty-first ICML}, page~79. ACM, 2004.

\bibitem{donmez2008paired}
Pinar Donmez and Jaime~G Carbonell.
\newblock Paired-sampling in density-sensitive active learning.
\newblock In {\em Proceedings of the 10th International Symposium on Artificial
  Intelligence and Mathematics}, 2008.

\bibitem{tomanek2009reducing}
Katrin Tomanek and Udo Hahn.
\newblock Reducing class imbalance during active learning for named entity
  annotation.
\newblock In {\em Proceedings of the fifth international conference on
  Knowledge capture}, pages 105--112. ACM, 2009.

\bibitem{joulin2017bag}
Armand Joulin, Edouard Grave, Piotr Bojanowski, and Tomas Mikolov.
\newblock Bag of tricks for efficient text classification.
\newblock In {\em Proceedings of the 15th EACL: Volume 2, Short Papers}, pages
  427--431, 2017.

\bibitem{bojanowski2017enriching}
Piotr Bojanowski, Edouard Grave, Armand Joulin, and Tomas Mikolov.
\newblock Enriching word vectors with subword information.
\newblock {\em Transactions of the Association for Computational Linguistics},
  5:135--146, 2017.

\bibitem{ertekin2009learning}
Seyda Ertekin.
\newblock {\em Learning in extreme conditions: Online and active learning with
  massive, imbalanced and noisy data}.
\newblock PhD thesis, The Pennsylvania State University, 2009.

\bibitem{attenberg2010unified}
Josh Attenberg, Prem Melville, and Foster Provost.
\newblock A unified approach to active dual supervision for labeling features
  and examples.
\newblock In {\em ECML/PKDD}, pages 40--55. Springer, 2010.

\bibitem{sindhwani2009uncertainty}
Vikas Sindhwani, Prem Melville, and Richard~D Lawrence.
\newblock Uncertainty sampling and transductive experimental design for active
  dual supervision.
\newblock In {\em Proceedings of the 26th ICML}, pages 953--960. ACM, 2009.

\bibitem{khudabukhsh2015building}
Ashiqur~R KhudaBukhsh, Paul~N Bennett, and Ryen~W White.
\newblock Building effective query classifiers: a case study in self-harm
  intent detection.
\newblock In {\em Proceedings of the 24th ACM CIKM conference}, pages
  1735--1738. ACM, 2015.

\bibitem{xChangeOrg}
XChange.org.
\newblock The rohingya survey 2017, 2017.
\newblock [Online; accessed 12-May-2019].

\bibitem{bhatia2018rohingya}
Abhishek Bhatia, Ayesha Mahmud, Arlan Fuller, Rebecca Shin, Azad Rahman, Tanvir
  Shatil, Mahmuda Sultana, KA~M Morshed, Jennifer Leaning, and Satchit Balsari.
\newblock The rohingya in cox’s bazar: When the stateless seek refuge.
\newblock {\em Health and human rights}, 20(2):105, 2018.

\bibitem{milton2017trapped}
Abul Milton, Mijanur Rahman, Sumaira Hussain, Charulata Jindal, Sushmita
  Choudhury, Shahnaz Akter, Shahana Ferdousi, Tafzila Mouly, John Hall, and
  Jimmy Efird.
\newblock Trapped in statelessness: Rohingya refugees in bangladesh.
\newblock {\em International journal of environmental research and public
  health}, 14(8):942, 2017.

\bibitem{lynch2014syria}
Marc Lynch, Deen Freelon, and Sean Aday.
\newblock {\em Syria's socially mediated civil war}.
\newblock Universit{\"a}ts-und Landesbibliothek Sachsen-Anhalt, 2014.

\bibitem{o2014online}
Derek O'Callaghan, Nico Prucha, Derek Greene, Maura Conway, Joe Carthy, and
  P{\'a}draig Cunningham.
\newblock Online social media in the syria conflict: Encompassing the extremes
  and the in-betweens.
\newblock In {\em Proceedings of the 2014 IEEE/ACM International Conference on
  Advances in Social Networks Analysis and Mining}, pages 409--416. IEEE Press,
  2014.

\bibitem{chowdhury2018sentiment}
Hemayet~Ahmed Chowdhury, Tanvir~Alam Nibir, Md~Islam, et~al.
\newblock Sentiment analysis of comments on rohingya movement with support
  vector machine.
\newblock {\em arXiv preprint arXiv:1803.08790}, 2018.

\bibitem{IndPak}
Shriphani Palakodety, Ashiqur~R. KhudaBukhsh, and Jaime~G. Carbonell.
\newblock Kashmir: {A} computational analysis of the voice of peace.
\newblock {\em CoRR}, abs/1909.12940, 2019.

\bibitem{haddal2009refugee}
Chad~C Haddal.
\newblock Refugee and asylum-seeker inflows in the united states and other oecd
  member states.
\newblock Congressional Research Service, Library of Congress, 2009.

\bibitem{blei2003latent}
David~M Blei, Andrew~Y Ng, and Michael~I Jordan.
\newblock Latent dirichlet allocation.
\newblock {\em JMLR}, 3(Jan):993--1022, 2003.

\bibitem{sievert2014ldavis}
Carson Sievert and Kenneth Shirley.
\newblock Ldavis: A method for visualizing and interpreting topics.
\newblock In {\em Proceedings of the workshop on interactive language learning,
  visualization, and interfaces}, pages 63--70, 2014.

\bibitem{o2010tweets}
Brendan O'Connor, Ramnath Balasubramanyan, Bryan~R Routledge, and Noah~A Smith.
\newblock From tweets to polls: Linking text sentiment to public opinion time
  series.
\newblock In {\em Fourth International AAAI Conference on Weblogs and Social
  Media}, 2010.

\bibitem{velikovich2010viability}
Leonid Velikovich, Sasha Blair-Goldensohn, Kerry Hannan, and Ryan McDonald.
\newblock The viability of web-derived polarity lexicons.
\newblock In {\em NAACL}, pages 777--785. Association for Computational
  Linguistics, 2010.

\bibitem{hamilton2016inducing}
William~L Hamilton, Kevin Clark, Jure Leskovec, and Dan Jurafsky.
\newblock Inducing domain-specific sentiment lexicons from unlabeled corpora.
\newblock In {\em EMNLP}, volume 2016, page 595. NIH Public Access, 2016.

\bibitem{settles2009active}
Burr Settles.
\newblock Active learning literature survey.
\newblock Technical report, University of Wisconsin-Madison Department of
  Computer Sciences, 2009.

\bibitem{pgj2017unsup}
Matteo Pagliardini, Prakhar Gupta, and Martin Jaggi.
\newblock {Unsupervised Learning of Sentence Embeddings using Compositional
  n-Gram Features}.
\newblock In {\em NAACL 2018}, 2018.

\bibitem{demszky2019analyzing}
Dorottya Demszky, Nikhil Garg, Rob Voigt, James Zou, Matthew Gentzkow, Jesse
  Shapiro, and Dan Jurafsky.
\newblock Analyzing polarization in social media: Method and application to
  tweets on 21 mass shootings.
\newblock In {\em Proceedings of the 17th Annual NAACL)}, 2019.

\bibitem{tong2001support}
Simon Tong and Daphne Koller.
\newblock Support vector machine active learning with applications to text
  classification.
\newblock {\em JMLR}, 2(Nov):45--66, 2001.

\end{thebibliography}

\end{document}